%% file: draft_v6.tex
\documentclass[conference]{IEEEtran}
\usepackage[
  top=0.80in,
  bottom=1.04in,      
  left=0.667in,       
  right=0.667in
]{geometry}
\IEEEoverridecommandlockouts
\usepackage{cite}
\usepackage{amsmath,amssymb,amsfonts}
\usepackage{algorithm}
\usepackage{algpseudocode}
\usepackage{graphicx}
\usepackage{textcomp}
\usepackage{xcolor}
\usepackage{cuted}
\usepackage{microtype}
\usepackage{svg}
\def\BibTeX{{\rm B\kern-.05em{\sc i\kern-.025em b}\kern-.08em
    T\kern-.1667em\lower.7ex\hbox{E}\kern-.125emX}}
\begin{document}
\include{commands}

\title{Interference-Aware Multiuser Hybrid Precoding for Coexistence with LEO Satellite Communication\\

}
\author{\IEEEauthorblockN{Nima Razavi, Murat Bayraktar, Nuria González Prelcic, Robert W. Heath Jr.}
\IEEEauthorblockA{\text{Dept. Electrical and Computer Engineering, University of California, San Diego, USA} \\
\{nrazavi, mbayraktar, ngprelcic, rwheathjr\} @ucsd.edu}}

\maketitle

\begin{abstract}
Interference from terrestrial networks can reduce the communication rate for low Earth orbit (LEO) satellites in the upper mid-band. To coexist in frequency, MIMO precoding can be used to reduce the signal that impinges on the LEO satellite. We present a beamforming algorithm designed for the hybrid architecture that incorporates a satellite interference penalty while optimizing the analog and digital precoders. Our algorithm optimizes the precoding at the base station (BS) within the set of precoders that null the interference to the satellite. Simulations demonstrate that our algorithm reduces the interference at the satellites and lowers the probability of violating prescribed LEO satellite protection thresholds, outperforming prior hybrid nulling algorithms. Results indicate that the algorithm maintains sum-rate within 3\% of the existing hybrid solutions, while effectively improving interference to noise power by 22.4 dB.
\end{abstract}
% \begin{IEEEkeywords}
% Spectrum sharing, spectrum coexistence, Upper Mid-Band, MIMO 
% % component, formatting, style, styling, insert
% \end{IEEEkeywords}
\section{Introduction}
The upper mid-band, or Frequency Range 3 (FR3), from 7-24 GHz offers a favorable balance of coverage and capacity trade-offs for next-generation systems \cite{Uppermidb2024}. The proposed spectrum range for FR3, however, is already occupied by various incumbent users. For instance, commercial low Earth orbit (LEO) satellite services operate at frequencies ranging from 10.7 to 12.7 GHz for the downlink (DL) and 14.0 to 14.5 GHz for the uplink (UL)\cite{CuiZhangPolUpperMidOppTechChal}. Changing the frequency assignment of satellite transceivers and sensing instruments is not a feasible solution to cellular network expansion into FR3, but spectrum coexistence opportunities are available \cite{sharingservices}. As non‐primary terrestrial operators seek to deploy in the upper mid-band, ensuring spectrum coexistence with these primary satellite incumbents is a regulatory and engineering challenge.

Carefully designed strategies at the BS can reduce interference between terrestrial infrastructure and satellites \cite{ortiz2024emergingngsoconstellationsspectral,exclusionzones,RahKisAloCoexistencemmWave2023, TerrSatNull2024}. One subset of methods to prevent satellite interference involves the creation of an exclusion zone, where BS transmit powers are limited to prevent interference to the satellite. Exclusion zone implementations, however, require additional overhead structure \cite{ortiz2024emergingngsoconstellationsspectral}, collaboration between BSs \cite{exclusionzones}, and access to spectrum access databases between each BS and satellite \cite{RahKisAloCoexistencemmWave2023}, limiting the available rate for satellite and terrestrial users.

MIMO beamforming provides a dimension that can be used to reduce interference. Relying on spatial nulling in the direction of the satellite, beamforming with a fully digital MIMO architecture can reduce BS interference to LEO satellites \cite{TerrSatNull2024}. BSs operating at the upper mid-band, however, incorporate a hybrid MIMO architecture \cite{introHybrid} to realize more antenna elements available on an array, as shown by the Giga-MIMO testbed created by Qualcomm \cite{Smee_2024}. The hybrid MIMO architecture, however, provides hardware limitations that complicate the interference nulling process. More work is needed to develop algorithms to ensure spectrum coexistence in various use cases. 

In this paper, we propose an algorithm for BSs using the hybrid MIMO architecture to mitigate their interference to LEO satellites while the BS communicates with DL user equipment (UE). Our proposed solution is a block coordinate descent (BCD) alternating optimization algorithm, which alternates between the analog and digital precoder \cite{Bertsekas_1999}. The algorithm relaxes hybrid precoder constraints to create a cost function and alternates between optimizing the analog and digital precoders. Within the proposed BCD algorithm, we construct a cost function that incorporates a satellite interference penalty. With the proposed cost function, we incorporate projected gradient descent (PGD) in the optimization algorithm using the closed-form gradients of the cost function. We evaluate the resulting average interference-to-noise power (INR) experienced at the satellites and sum-rate of the UEs to compare their performance with previous hybrid interference nulling strategies.

\emph{Notation:} Let $\mathbb{C}$ denotes the field of complex numbers. Let $(\cdot)^\top$, $(\cdot)^{*}$, $\|\cdot\|$ and $\|\cdot\|_\rmF$ indicate the transpose, conjugate transpose, Euclidean norm, and Frobenius norm of a vector or matrix, respectively. We denote a capital, bolded letter $\mathbf{A}$ as a matrix, and a lowercase, bolded letter $\mathbf{a}$ as a vector. We denote $\mathbf{A}[i,j]$ as the $i$-th row and $j$-column of matrix $\mathbf{A}$. We denote $\{\mathbf{w}_{i}\}_{i=1}^N$ as a set of N vectors. Let $\mathcal{CN}(\mu, \sigma^2)$ be the complex normal circularly symmetric distribution with mean $\mu$ and variance $\sigma^2$. We denote $\ek$ as the basis vector of a specific size with 1 at index $\u$ and 0 elsewhere.
\section{System model}
We consider the DL transmission of a terrestrial cellular network operating at an upper mid-band frequency using a uniform rectangular array (URA) with $\NT$ total antennas. The BS transmits simultaneously to $\K$ multi-antenna UEs, each with an $\NR$-element uniform linear array (ULA), each receiving a single stream. At the same time as the BS DL transmission, a LEO satellite operating at the same frequency receives the UL transmission from a satellite ground user. 

\begin{figure}[h!]
    \centering
\includegraphics[width=0.42\textwidth]{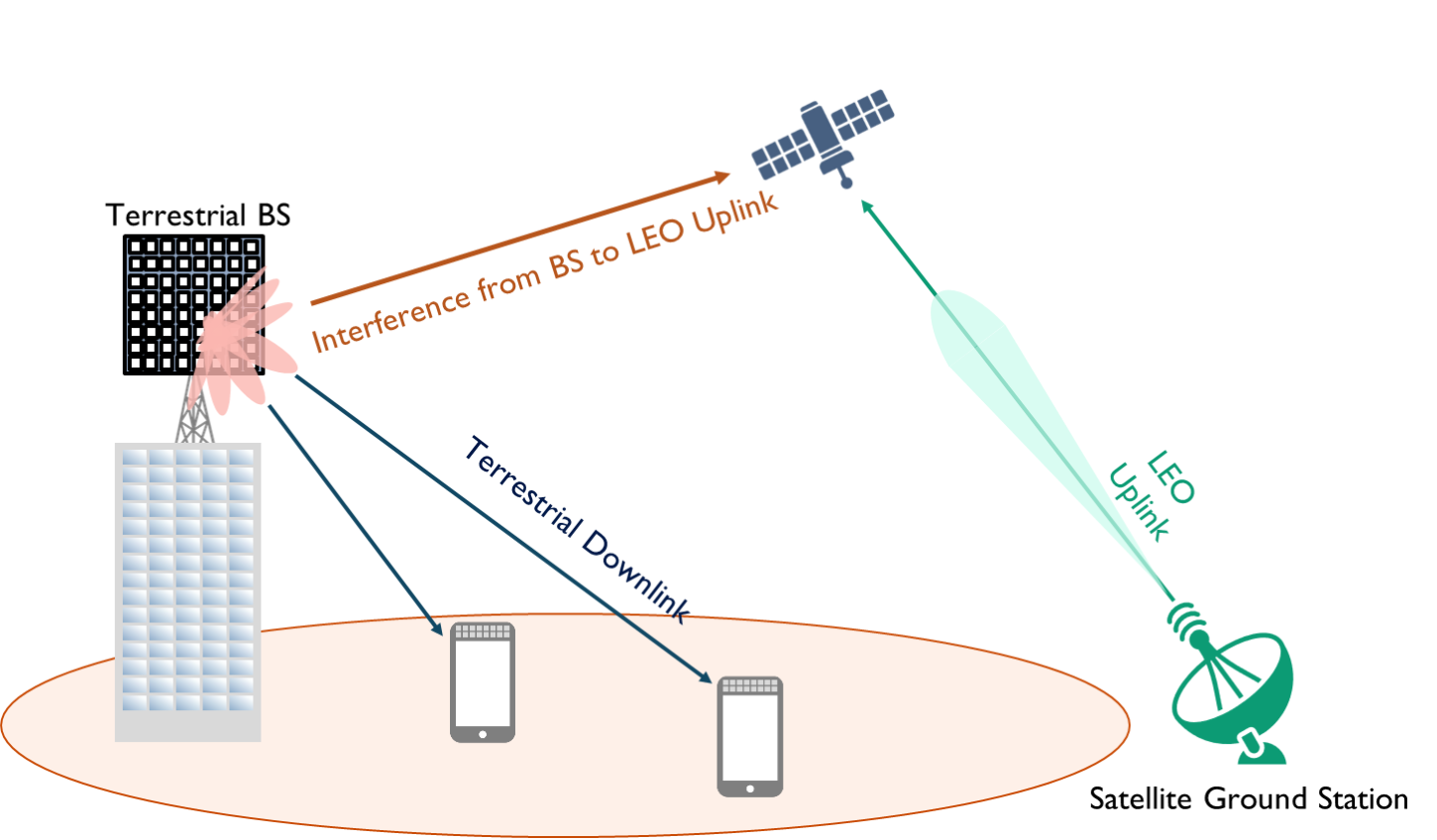}
\caption{Illustration of the coexistence model with a terrestrial BS, DL UEs, LEO satellites, and a satellite ground station}
\label{model}
\end{figure}

We consider a BS equipped with the hybrid MIMO architecture with $\NRF$ fully connected radio frequency (RF) chains and multiple UEs equipped with fully digital MIMO antennas. We provide an illustration of the system model in Fig. \ref{model}.
\vspace{-12pt}
\subsection{Channel}
We define the various channels in the system model. We denote the channel between the BS and satellite $i$, one of $\N$ total LEO satellites, as  \(\hSat\in\mathbb{C}^{\NT\times1}\). We simplify the system by limiting both channels to the narrowband channel models. We consider  $\alphal \in \mathbb{C}$ as the complex gain (including pathloss), \(\thetal\), \(\phil\) as the angles of arrival/departure, and $\aT(\thetal)\in\mathbb{C}^{\NT\times1}$ and $\aR(\phil)\in \mathbb{C}^{\NR\times1}$ as the transmit and receive steering vectors. The narrowband DL channel from the BS to UE $u$ is modeled as the sum of $\Lu$ paths as
\begin{equation}
    \Hk = \sum_{\ell=1}^{\Lu} \alphal\, \aR(\thetal)\, \aT(\phil)^*.
\end{equation}
We assume the channel between the BS and the UE is known. We assume the channel between the BS and LEO satellite is the line-of-sight (LOS) channel because of the power predominance of the LOS path compared to the non-LOS paths in this system. The BS knows the position of the LEO satellite, collected from ephemeris data of the satellite's known trajectory. No assumption about the antenna geometry at the satellite is known. No coordination exists between the BS and the LEO satellite. We denote the channel between the BS and LEO satellite $i$ as $\hSat$. Using the known satellite position, we define the channel between the BS and LEO satellite $i$ as
\begin{equation}
\label{hSatChannel}
    \hSat = \aT(\tthetal,\tphil),
\end{equation}
 which is the array steering vector with azimuth $\tphil$, and elevation $\tthetal$. The LOS channel assumption is also used in constructing the satellite channel in previous works \cite{TerrSatNull2024}.   
\subsection{Received signal}
We define the analog precoder as $\FRF\in\mathbb{C}^{{\NT}\times\NRF}$ with unit-amplitude entries such that $ |\FRF[m,n]|=1$ for all matrix indices $m,n$, considering the given hybrid architecture at the BS. We construct the digital precoder matrix as $\FBB\in\mathbb{C}^{\NRF\times \K},$ such that $\FBB = [\fBB,\dots \fBBK]$, where $\fBBx$ is the digital precoder for the $\u$-th UE. Here, the hybrid architecture with phase shifters is fully connected to the RF chains. We denote the hybrid precoding matrix as $\FHYB = \FRF\FBB$. The hybrid precoding matrix is constrained by the total transmit power available at the BS, which is denoted $\Pmax$. The transmit power constraint of the hybrid precoding matrix is therefore $\bigl\|\FRF\,\FBB\bigr\|_{\rmF}^2 \le \Pmax$. We denote the overall transmit signal from the BS as $\bs = [s_1,\dots,s_{\u}]^{\top}$. We assume each symbol $s_i$ is independent, and all symbols are chosen from the same distribution. Because of the fully digital MIMO architecture at each UE, we denote the digital combiner for UE $\u$ as $\wBB \in \mathbb{C}^{\NR\times1}$ which is normalized to $\| \wBB\|  = 1$. The noise vector is denoted by $\ndl = [z_{1,\u}\dots,z_{\NR,\u}]$ where $z_{r,\u}$ is the noise experience at receive antenna $r$ of UE $\u$. The distribution of the noise vector for UE $\u$ is $\ndl \sim \mathcal{CN}(0,\sigma_{\u} ^2\mathbf{I}_{\NR})$. The received signal at UE $\u$ after applying the fully digital combiner $\wBB$ is
\begin{equation}
    \ydlu = \wBB^*\Hk\FRF\FBB\mathbf{s} + \wBB^*\ndl.
\end{equation}
To determine the quality of $\FRF$, $\FBB$, and $\wBB$ in the considered system, we define two performance metrics.
\subsection{Satellite interference metric}
We define the first performance metric as the total interference experienced by the satellites from the BS. We construct $\HSat$ from the LOS channel at the BS to each satellite such that $\HSat = [\hSati,\dots,\hSatN]$ for the $\N$ satellite channels \eqref{hSatChannel}. We denote the satellite interference penalty as the total power of the transmit beamforming gain in the direction of each satellite from the resulting hybrid precoder. The interference penalty is denoted as
\begin{equation}
\label{inr}
    P_\text{interference}(\FRF,\FBB) = ||\HSat\FRF\FBB||^2_{\rmF}.
\end{equation}
\subsection{UE Sum-rate metric}
We define the second metric as the sum-rate of the terrestrial downlink UEs. Using $\ek$ of length $\K$, we define the signal-to-interference-plus-noise ratio (SINR) for UE $\u$, denoted as $\Sk$, and sum-rate expression for UE $\u$, denoted as $\Ru$, such that
\begin{multline}  \label{SINR}
\Sk(\FRF,\FBB,\wBB)
= \\
\frac{\bigl|\wBB^{*}\Hk\FRF\FBB\ek\bigr|^{2}}
     {\sigma_{\u}^{2} + \sum_{j\neq \u}\bigl|\wBB^{*}\Hk\FRF\FBB\ej\bigr|^{2}},
\end{multline}
\begin{multline}  \label{rate}
\Ru(\FRF,\FBB,\wBB)
= \\
\sum_{\u=1}^{\K} \log\!\bigl(1+\Sk(\FRF,\FBB,\wBB)\bigr) .
\end{multline}
The equation \eqref{SINR} represents the total rate of all users, accounting for inter-user interference. We do not incorporate interference from the satellite ground stations into the calculation of SINR because satellite ground stations have highly directional antennas that are aligned to face the sky \cite{yost2024state}. We also focused on the case where the satellite communication is in UL mode.
\section{Problem formulation}
Our goal is to design the precoders at the BS based on competing objectives. We aim to maximize UE sum-rate and to minimize interference at the satellite from the BS. We formulate a constrained optimization problem by incorporating the performance metrics, satellite interference \eqref{inr} and sum-rate \eqref{rate}, constrained by the unit-modulus entries of the analog precoder and the maximum transmit power. The goal of considering the constraints and performance metrics into one problem is to design $\FRF$,$\FBB$, and $\wBB$ for each UE to maximize the sum-rate of the UEs while minimizing interference from the BS to the LEO satellites.

We consider a penalty-based objective function to penalize an increase in satellite interference while maximizing UE sum-rate. Previous hybrid precoding designs use penalty method-based objective functions to minimize BS power budget \cite{ma2025digitalhybridprecodingdesigns}. To scale the penalty term of the satellite interference according to the maximum satellite interference desired, we use a penalty parameter $\lambdaSat$. We express this problem as 
\begin{subequations}\label{objective}
\begin{align}
  \max_{\FRF,\,\FBB,\,\{\wBB\}_{\u=1}^{\K}} \quad
  & \Ru(\FRF,\FBB,\wBB) \nonumber \\
  & {} - \lambda_{\text{Sat}}\,P_{\text{interference}}(\FRF,\FBB) \label{objective:cost} \\
  \text{s.t.} \quad \quad \quad \quad \:\:
  & \|\FRF\FBB\|_{\mathrm{F}}^{2} = \Pmax, \label{objective:a} \\
  & |\FRF(m,n)|^2 = 1,\quad \forall\,m,n. \label{objective:b}
\end{align}
\end{subequations}
We impose two constraints on the optimization problem \eqref{objective}: the total transmit power is the maximum possible transmit power $\Pmax$ \eqref{objective:a}, and every entry of the RF precoder has magnitude one \eqref{objective:b}. These constraints are consistent with hybrid MIMO design problems studied previously \cite{hybridprecoding}.
\section{Joint sum–rate maximization and satellite interference nulling}
\begin{algorithm}
\caption{Gradient–based hybrid precoding with satellite nulling}
\label{alg}
\begin{algorithmic}[1]
\Require $\{\Hk\}_{\u = 1}^{\K},\HSat,\Pt,\lambdaSat,\alpha,\itBB,\itRF,\tot$
\State Initialize $\mathbf{F}_{\rm RF}^{(0)}$, $\mathbf{F}_{\rm BB}^{(0)}$
\For{$t=1$ {\bf to} $\tot$}
  \State \emph{Digital update:}
  \For{$n=1$ {\bf to} $\itBB$}
    \State Compute $\nabla_{\FBB}\cC$ by \eqref{grad_BB}
    \State $\FBB\,\gets\,\FBB-\alpha\,\nabla_{\FBB}\cC$
    \State Normalize $\FBB\,\gets\,\sqrt{\Pt}\,\frac{\FBB}{\|\FRF\FBB\|_{\rmF}}$
  \EndFor
  \State \emph{Analog update:}
  \For{$m=1$ {\bf to} $\itRF$}
    \State Compute $\nabla_{\FRF}\cC$ by \eqref{grad_RF}
    \State $\FRF\,\gets\,\exp\!\bigl(j\,\angle(\FRF-\alpha\,\nabla_{\FRF}\cC)\bigr)$
  \EndFor
  \State Update combiners $\{\wBB\} _{\u=1}^{\K}$
\EndFor
\Ensure $\FRF,\,\FBB$
\end{algorithmic}
\end{algorithm}
We describe the methods used for designing $\FRF$, $\FBB$, and $\wBB$, which attempt to maximize \eqref{objective}. We note that \eqref{objective} is non-convex because of the unit-modulus constraint on the analog precoder and that the analog and digital precoders are coupled. We also note that the combiner for each UE can be optimized separately. We separate the optimization of each argument of \eqref{objective} through a BCD algorithm. We describe separately combiner and precoder optimization, which are then integrated into Algorithm \ref{alg}.
\vspace{-6pt}
\subsection{Combiner optimization}
We define a method for optimizing the digital combiner $\wBB$ while fixing $\FRF$ and $\FBB$. We note that the $\log(1+\Sk)$ is a concave function and is monotonically increasing. Thus, maximizing $\log(1+\Sk)$ is equivalent to maximizing $\Sk$. Since we aim to maximize $\Sk$ through the design of $\wBB$, the UE solves for $\wBB$ through a Rayleigh Quotient problem \cite{haykin2002adaptive}. We denote
\begin{subequations}
    \begin{align}
        \Rsig & \propto \Hk\,\FRF\FBB\ek (\FRF\FBB\ek)^*\Hk^*,\\
        \Ryy & \propto \sigma_{\u}^2\bI + \sum_{j\neq \u}\Hk\,\FRF\FBB\ej (\FRF\FBB\ej)^*\,\Hk^*.
    \end{align}
\end{subequations}
as auxiliary variables for UE $\u$. The combiner $\wBB$ has a closed-form solution using the eigenvector corresponding to the maximum eigenvalue of $\Ryy^{-1}\Rsig$. 
\vspace{-2pt}
\subsection{Precoder optimization}
We incorporate projected gradient descent (PGD) \cite{Bertsekas_1999} to minimize a cost function derived from \eqref{objective}. PGD is efficient with the appropriate selection of a cost function because of the available closed-form gradient expression with respect to the analog and digital precoders. By considering the objective function \eqref{objective}, we form the following cost function to minimize:
\begin{multline}\label{eq:cost}
    \cC(\FRF,\FBB)
=-\sum_{\u=1}^\K\log\bigl(1+\Sk\bigr)
\; \\
+\;\lambdaSat\,\mathrm{tr}\bigl(\FBB^*\FRF^*\HSat^*\HSat\,\FRF\,\FBB\bigr).
\end{multline}
We express the closed-form gradient of \eqref{eq:cost} with respect to the analog precoder $\FRF$ and digital precoder $\FBB$. Because of a lack of space, we omit the derivation here. To shorten the equation, we define the following auxiliary variables
\begin{subequations}
\label{auxs}
    \begin{align}
        \uk &\triangleq \wBB^*\Hk\FRF\FBB\ek,\\
        \num &\triangleq |\uk|^2,\\
        \den &\triangleq \sigma_{\rm \u}^2+\sum_{j\neq \u}^{\K}|\wBB^*\Hk\FRF\FBB\ej|^2,\\
        \Mk & \triangleq \Hk^*\wBB\wBB^*\Hk,\\
        \mathbf{N}_{\u,j} & \triangleq \Mk\FRF\FBB\ek(\FBB\ej)^*.
    \end{align}
\end{subequations}
We incorporate the auxiliary variables \eqref{auxs} into the definitions of the closed-form gradients for the analog and digital processor, which are derived as
\begin{multline}
\label{grad_RF}
\nabla_{\FRF}\cC
=2\lambdaSat\,\HSat^*\HSat\,\FRF\,\FBB\FBB^*
\\
+\sum_{\u=1}^{\K} 
\Bigl[
\tfrac{2}{\den + \num}\Bigl(\mathbf{N}_{\u,\u} +  \mathbf{N}_{\u,\u}^*\Bigr) \\
-\tfrac{2\num}{\den(\den+\num)}
\sum_{j \neq \u}^\K \Bigl(\mathbf{N}_{\u,j} + \mathbf{N}_{\u,j}^*\Bigr)
\Bigr],
\end{multline}
\begin{multline}
\label{grad_BB}
\nabla_{\FBB}\cC
=2\lambdaSat\,\FRF^*\HSat^*\HSat\FRF\,\FBB\\
\quad
+\sum_{\u=1}^{\K} 
\Bigl[
  \tfrac{2}{\den+\num}\FRF^*\Mk\FRF\FBB\ek\ek^{\top}\\
  -\tfrac{2\num}{\den(\den+\num)}
   \sum_{j \neq \u}^{\K} \FRF^*\Mk\FRF\FBB\ej\ej^{\top}
\Bigr].
\end{multline}
We denote a fixed step size for PGD, common for both updates, as $\alpha$. We use a fixed step size because step size optimization, such as line search and backtracking, involves additional per-iteration processing, increasing the overall complexity and run-time. A common practice is to use pre-defined hand-tuned constant step sizes. Learning or manually tuning the fixed step size before can help to rapidly optimize precoders \cite{stepsize}. We incorporate design decisions to keep the complexity low because baseline hybrid interference nulling methods are of lower complexity compared to iterative methods. We apply the gradient step to each precoder using the closed-form gradient. We then project each precoder back into the feasible space to satisfy the constraints expressed in \eqref{objective}. For the unit norm constraint of the analog precoder, we apply PGD on $\FRF$ during analog precoder update, then project each entry of $\FRF$ onto the complex unit circle. For the transmit power constraint, we apply PGD on $\FBB$ during digital precoder update, then normalize $\FBB$ such that the power of the total hybrid matrix matches $\Pt$. As expressed in \eqref{objective}, we do not consider power control in our solution. The total transmit power is set to the maximum available power $\Pt = \Pmax$.
\subsection{BCD Algorithm}
We design a BCD algorithm to optimize both precoders using PGD and the closed-form gradient expressions. BCD is a natural choice for optimizing the coupled variables $\FRF$ and $\FBB$. The proposed BCD Algorithm \ref{alg} incorporates PGD on both analog and digital precoders by performing a fixed number of iterations of gradient steps for the digital precoder denoted by $\itBB$, then optimizing for a fixed number of iterations of gradient steps for the analog precoder, denoted by $\itRF$. BCD using fixed analog and digital update iterations not only performs close to high-performance precoders that use alternating minimization with manifold optimization, but also involves smaller complexity \cite{BCD}. Finally, $\tot$ denotes the total number of outer loop iterations.
\section{Performance Evaluation}
We assess the efficacy of our proposed hybrid precoding scheme by comparing its DL sum‐rate and satellite interference to noise power (INR) performance against several benchmark interference‐nulling methods. All simulations assume a BS equipped with an $8\times 8$ URA, $\NRF = 8$ RF chains, and $\K=2$ UEs, each with $\NR = 2$ antennas. To test coexistence, we select $\N=2$ LEO satellites to null. The LEO satellite locations were chosen from known trajectories with corresponding atmospheric attenuation. In all simulations and noise calculations, we use a carrier frequency of 14 GHz and a bandwidth of 200 MHz.   
\begin{figure}[h!]
    \centering
\includegraphics[width=0.45\textwidth]{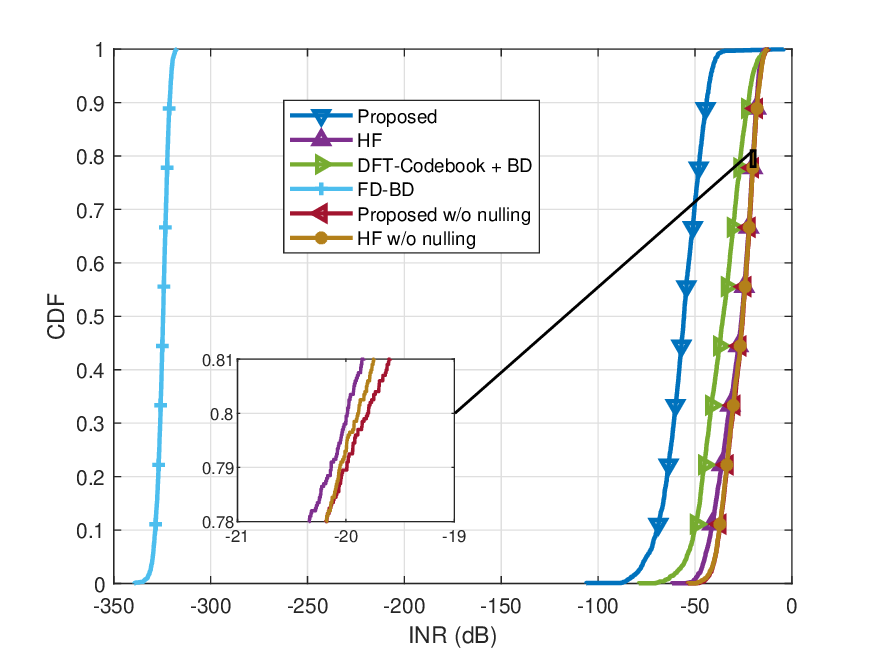}
\vspace{-4pt}
\caption{CDF of satellite INR for 2 UE, 8x8 BS array with 8 RF chains and 2 satellites present with $\lambdaSat=10$ for the proposed solution. As seen in the figure, the proposed method reduces the INR compared to all baseline hybrid solutions, and INR is only surpassed by a fully digital MIMO architecture.}
\label{inr_fig}
\end{figure}
\begin{figure}[h!]
    \centering
\includegraphics[width=0.45\textwidth]{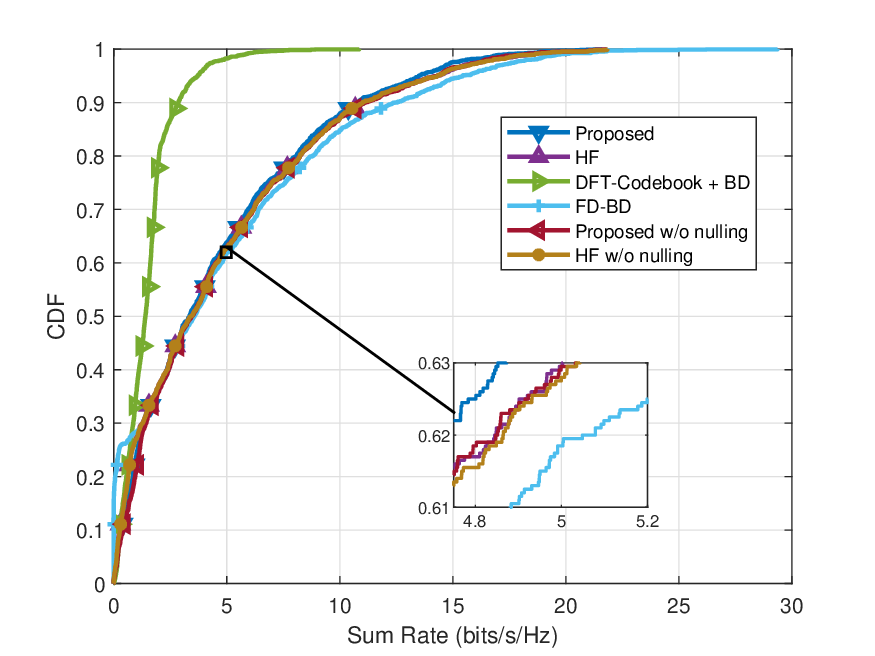}
\vspace{-4pt}
\caption{CDF of UE Sum-Rate for 2 UEs, 8x8 BS array with 8 RF chains and 2 satellites present with $\lambdaSat=10$ for the proposed solution. The figure illustrates the ability of the proposed solution to achieve sum-rate near other hybrid solutions and the fully digital architecture.}
\label{sumr_fig}
\end{figure}
\begin{figure}[h!]
    \centering
\includegraphics[width=0.45\textwidth]{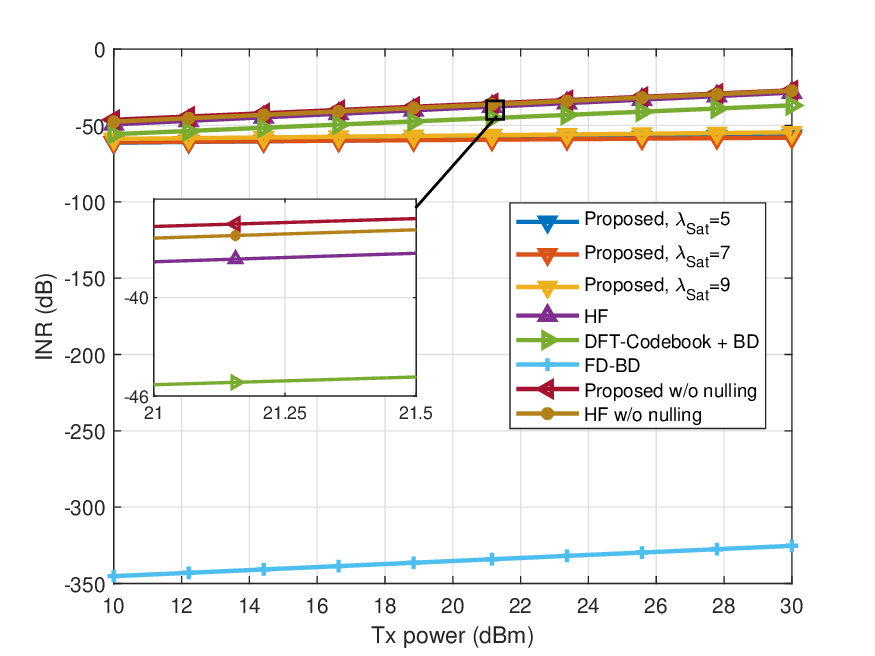}
\vspace{-4pt}
\caption{BS array transmit power and the resulting satellite INR. Across various transmit powers, the proposed method with each selection of $\lambdaSat$ improves the satellite INR.}
\label{tx_vs_inr}
\end{figure}
\begin{figure}[h!]
    \centering
\includegraphics[width=0.45\textwidth]{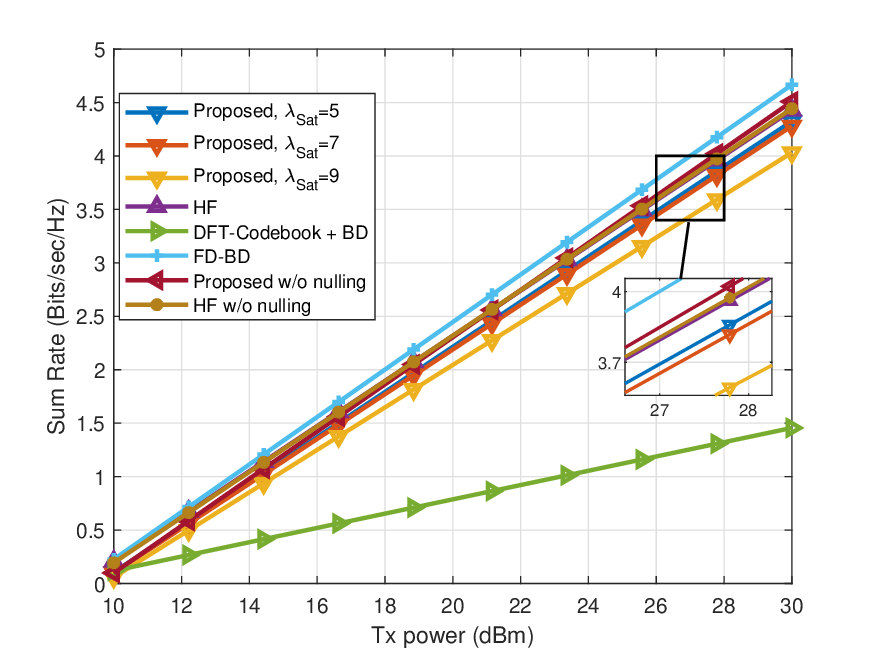}
\vspace{-4pt}
\caption{BS Array transmit power and the resulting UE sum-rate. Across various transmit powers, the proposed method with each selection of $\lambdaSat$ attains a similar sum-rate to other hybrid solutions.}
\label{tx_vs_sumr}
\end{figure}
\subsection{Benchmark precoding methods}
We define several baseline methods to compare the performances.
\begin{enumerate}[]
\item \textbf{Fully digital block diagonalization (FD‐BD):}  
    We incorporate classical Block Diagonalization (BD) \cite{blockdiag} with a fully digital array (no RF‐chain or phase shifter constraints) as an upper bound. To design the BD-based precoder for each UE, we stack the other UE channels together, append the satellite channels, and compute the null-space, yielding a precoder that nulls interference to the other UEs and satellites at the same time. This results in the highest achievable sum‐rate under perfect interference cancellation.
    \item \textbf{Hybrid factorization (HF).}  
    We compute the fully digital precoder via BD to eliminate multi-user interference. Using the resulting FD-BD matrix, we extract the first $\NRF$ dominant columns of this fully digital matrix, normalize each entry to satisfy the unit-modulus constraint of the analog precoder matrix, and apply a least-squares pseudoinverse to derive the corresponding digital precoder. This yields a low-complexity hybrid approximation of the BD solution.
    \item \textbf{DFT‐codebook + BD.}  
    We select analog precoding vectors from a predefined Discrete Fourier Transform (DFT) codebook; for each UE, we select the beam index that maximizes its array gain. Once the $\NRF$ beams are fixed, we apply BD in the reduced-dimensional subspace to compute the digital precoder, nulling inter‐UE interference among selected beams.
    \item \textbf{Gradient based without nulling.}  
    We optimize the hybrid precoders with the proposed projected gradient algorithm when we do not attempt to penalize the satellite interference, which is accomplished by setting $\lambdaSat = 0$.
    \item \textbf{HF without nulling.}  
    We compute the precoders using HF without concatenating the satellite channel in BD.
\end{enumerate}
We keep all optimization parameters constant to test Algorithm \ref{alg}. We set the step size as $\alpha = 10^{-4}$, the inner digital and analog update steps as  $\itBB = \itRF = 5$ respectively, and the total number of outer loop iterations as $\tot = 20$.

We use the HF method to initialize $\FRF$ and $\FBB$. We initialize the gradient based without nulling baseline using HF without nulling.
\begin{figure*}[t!]
    \centering
\includegraphics[width=0.80\textwidth]{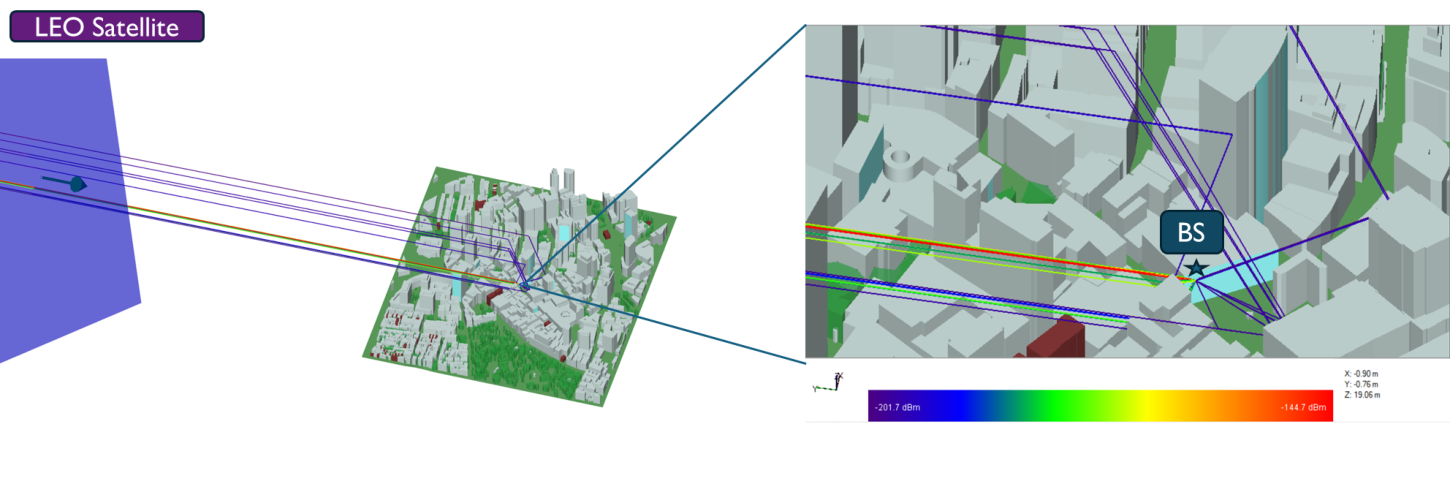}
\vspace{-30pt}
\caption{Example Wireless Insite channel simulation from 1 LEO Satellite to a fixed BS located 20 meters above ground on the side of a building in a dense urban environment. The star denotes the BS located on the top of a building with rays propagating to the blue surface of the LEO satellite.}
\label{wirelessinsite}
\end{figure*}
\subsection{Channel and scenario modeling}
We generate realistic terrestrial channels using Wireless Insite \cite{remcomWirelessPropagation}. A Wireless Insite channel simulation illustrating propagation paths from the satellite and BS in the urban environment is shown in Fig. \ref{wirelessinsite}. We place the BS  on top of a building within the urban environment with a height of 20 m. We selected UE locations from outdoor vehicle locations within the urban environment on the surrounding ground plane. We simulate propagation within a dense urban environment using glass and concrete material for buildings. We use a known LEO satellite trajectory that orbits at an altitude of 100 km. We incorporate the instantaneous azimuth/elevation angles at the BS because of the known trajectory data.  We also include atmospheric attenuation as part of the path‐loss model for the BS to satellite path.
\subsection{Numerical results}
To quantify performance, we compute the sum-rate using \eqref{SINR} and the average INR across all satellites. Using large-scale pathloss and noise power at satellite $i$ denoted $L_i$ and $\sigma_{i}^2$ respectively, INR is calculated in dB as
  \[
    \mathrm{INR}_i
    =10\log_{10}\!\Bigl(\tfrac{\Pt|\hSat^* \FRF\FBB\,|^2}{{L}_{i}\sigma_{i}^2}\Bigr).
  \]
The cumulative density functions (CDFs) of the INR and the DL sum-rate are shown in Fig. \ref{inr_fig} and Fig. \ref{sumr_fig}, respectively. Our proposed nulling algorithm achieves within 2.69\% of the average HF rate (the highest rate of the hybrid baseline methods) while reducing the mean satellite INR by 22.4 dB compared to the DFT‐Codebook + BD baseline (the lowest INR of the hybrid architecture baselines). The FD-BD method beats the hybrid solutions as expected.
 
We use the Interference Protection Criteria guidelines \cite{paul2005interference} to recognize if the proposed nulling scheme contributes to reducing harmful interference at the satellites. Table 4-7 of the guidelines notes that all non-primary sources, such as cellular networks in the upper mid-band, must maintain an INR below -20 dB at all times. The DFT+BD hybrid benchmark exceeds -20 dB in 7.70\% of trials, and the HF benchmark exceeds this in 20.1\% of trials, while the proposed gradient-based method that incorporates nulling only exceeds -20 dB INR in 0.05\% of trials.

We further demonstrate the average sum-rate and INR performance across several transmit powers, linearly increasing from 0.01 to 1 W (10 to 30 dBm) in Fig. \ref{tx_vs_inr} and \ref{tx_vs_sumr}. We evaluate several values of the satellite interference penalty parameter $\lambda_{\rm Sat}$. Figures \ref{tx_vs_inr} and \ref{tx_vs_sumr} demonstrate the ability of the proposed solution to outperform baselines across the range of BS array powers. The resulting INRs of the proposed method for different values of $\lambdaSat$ are all below the INR threshold, with a slight loss of spectral efficiency as $\lambdaSat$ is increased.
\section{Conclusion}
In this paper, we proposed a hybrid precoding algorithm that incorporates interference from cellular networks to LEO satellite UL. We devised a block coordinate descent algorithm with projected gradient that penalizes satellite interference while maximizing sum-rate. Compared to hybrid nulling benchmarks, simulation results confirm the proposed method's ability to decrease interference the furthest below the interference protection criterion while maintaining an adequate rate.
\section{Acknowledgments}
This material was based upon work supported by the UCSD ECE Departmental Fellowship.
\bibliographystyle{IEEEtran} % Use an appropriate bibliography style
\bibliography{MyReferences} % Include your .bib file

% \vspace{12pt}

\end{document}

%% file: commands.tex
% blackboard lowercase
\def\bydef{:=}
\def\bba{{\mathbb{a}}}
\def\bbb{{\mathbb{b}}}
\def\bbc{{\mathbb{c}}}
\def\bbd{{\mathbb{d}}}
\def\bbee{{\mathbb{e}}}
\def\bbff{{\mathbb{f}}}
\def\bbg{{\mathbb{g}}}
\def\bbh{{\mathbb{h}}}
\def\bbi{{\mathbb{i}}}
\def\bbj{{\mathbb{j}}}
\def\bbk{{\mathbb{k}}}
\def\bbl{{\mathbb{l}}}
\def\bbm{{\mathbb{m}}}
\def\bbn{{\mathbb{n}}}
\def\bbo{{\mathbb{o}}}
\def\bbp{{\mathbb{p}}}
\def\bbq{{\mathbb{q}}}
\def\bbr{{\mathbb{r}}}
\def\bbs{{\mathbb{s}}}
\def\bbt{{\mathbb{t}}}
\def\bbu{{\mathbb{u}}}
\def\bbv{{\mathbb{v}}}
\def\bbw{{\mathbb{w}}}
\def\bbx{{\mathbb{x}}}
\def\bby{{\mathbb{y}}}
\def\bbz{{\mathbb{z}}}
\def\bb0{{\mathbb{0}}}
\def\bbone{{\mathbb{1}}}

% Bold lowercase
\def\bydef{:=}
\def\ba{{\mathbf{a}}}
\def\bb{{\mathbf{b}}}
\def\bc{{\mathbf{c}}}
\def\bd{{\mathbf{d}}}
\def\bee{{\mathbf{e}}}
\def\bff{{\mathbf{f}}}
\def\bg{{\mathbf{g}}}
\def\bh{{\mathbf{h}}}
\def\bi{{\mathbf{i}}}
\def\bj{{\mathbf{j}}}
\def\bk{{\mathbf{k}}}
\def\bl{{\mathbf{l}}}
\def\bmm{{\mathbf{m}}} % changed to avoid conflict with bold
\def\bn{{\mathbf{n}}}
\def\bo{{\mathbf{o}}}
\def\bp{{\mathbf{p}}}
\def\bq{{\mathbf{q}}}
\def\br{{\mathbf{r}}}
\def\bs{{\mathbf{s}}}
\def\bt{{\mathbf{t}}}
\def\bu{{\mathbf{u}}}
\def\bv{{\mathbf{v}}}
\def\bw{{\mathbf{w}}}
\def\bx{{\mathbf{x}}}
\def\by{{\mathbf{y}}}
\def\bz{{\mathbf{z}}}
\def\bzero{{\mathbf{0}}}
\def\bone{{\mathbf{1}}}

% Bold capital letters
\def\bA{{\mathbf{A}}}
\def\bB{{\mathbf{B}}}
\def\bC{{\mathbf{C}}}
\def\bD{{\mathbf{D}}}
\def\bE{{\mathbf{E}}}
\def\bF{{\mathbf{F}}}
\def\bG{{\mathbf{G}}}
\def\bH{{\mathbf{H}}}
\def\bI{{\mathbf{I}}}
\def\bJ{{\mathbf{J}}}
\def\bK{{\mathbf{K}}}
\def\bL{{\mathbf{L}}}
\def\bM{{\mathbf{M}}}
\def\bN{{\mathbf{N}}}
\def\bO{{\mathbf{O}}}
\def\bP{{\mathbf{P}}}
\def\bQ{{\mathbf{Q}}}
\def\bR{{\mathbf{R}}}
\def\bS{{\mathbf{S}}}
\def\bT{{\mathbf{T}}}
\def\bU{{\mathbf{U}}}
\def\bV{{\mathbf{V}}}
\def\bW{{\mathbf{W}}}
\def\bX{{\mathbf{X}}}
\def\bY{{\mathbf{Y}}}
\def\bZ{{\mathbf{Z}}}

% Blackboard capital letters
\def\bbA{{\mathbb{A}}}
\def\bbB{{\mathbb{B}}}
\def\bbC{{\mathbb{C}}}
\def\bbD{{\mathbb{D}}}
\def\bbE{{\mathbb{E}}}
\def\bbF{{\mathbb{F}}}
\def\bbG{{\mathbb{G}}}
\def\bbH{{\mathbb{H}}}
\def\bbI{{\mathbb{I}}}
\def\bbJ{{\mathbb{J}}}
\def\bbK{{\mathbb{K}}}
\def\bbL{{\mathbb{L}}}
\def\bbM{{\mathbb{M}}}
\def\bbN{{\mathbb{N}}}
\def\bbO{{\mathbb{O}}}
\def\bbP{{\mathbb{P}}}
\def\bbQ{{\mathbb{Q}}}
\def\bbR{{\mathbb{R}}}
\def\bbS{{\mathbb{S}}}
\def\bbT{{\mathbb{T}}}
\def\bbU{{\mathbb{U}}}
\def\bbV{{\mathbb{V}}}
\def\bbW{{\mathbb{W}}}
\def\bbX{{\mathbb{X}}}
\def\bbY{{\mathbb{Y}}}
\def\bbZ{{\mathbb{Z}}}

% Calligraphic capital letters
\def\cA{\mathcal{A}}
\def\cB{\mathcal{B}}
\def\cC{\mathcal{C}}
\def\cD{\mathcal{D}}
\def\cE{\mathcal{E}}
\def\cF{\mathcal{F}}
\def\cG{\mathcal{G}}
\def\cH{\mathcal{H}}
\def\cI{\mathcal{I}}
\def\cJ{\mathcal{J}}
\def\cK{\mathcal{K}}
\def\cL{\mathcal{L}}
\def\cM{\mathcal{M}}
\def\cN{\mathcal{N}}
\def\cO{\mathcal{O}}
\def\cP{\mathcal{P}}
\def\cQ{\mathcal{Q}}
\def\cR{\mathcal{R}}
\def\cS{\mathcal{S}}
\def\cT{\mathcal{T}}
\def\cU{\mathcal{U}}
\def\cV{\mathcal{V}}
\def\cW{\mathcal{W}}
\def\cX{\mathcal{X}}
\def\cY{\mathcal{Y}}
\def\cZ{\mathcal{Z}}

% Sans serif capital letters
\def\sfA{\mathsf{A}}
\def\sfB{\mathsf{B}}
\def\sfC{\mathsf{C}}
\def\sfD{\mathsf{D}}
\def\sfE{\mathsf{E}}
\def\sfF{\mathsf{F}}
\def\sfG{\mathsf{G}}
\def\sfH{\mathsf{H}}
\def\sfI{\mathsf{I}}
\def\sfJ{\mathsf{J}}
\def\sfK{\mathsf{K}}
\def\sfL{\mathsf{L}}
\def\sfM{\mathsf{M}}
\def\sfN{\mathsf{N}}
\def\sfO{\mathsf{O}}
\def\sfP{\mathsf{P}}
\def\sfQ{\mathsf{Q}}
\def\sfR{\mathsf{R}}
\def\sfS{\mathsf{S}}
\def\sfT{\mathsf{T}}
\def\sfU{\mathsf{U}}
\def\sfV{\mathsf{V}}
\def\sfW{\mathsf{W}}
\def\sfX{\mathsf{X}}
\def\sfY{\mathsf{Y}}
\def\sfZ{\mathsf{Z}}

% Bold sans serif capital letters
\def\bsfA{\bm{\mathsf{A}}}
\def\bsfB{\bm{\mathsf{B}}}
\def\bsfC{\bm{\mathsf{C}}}
\def\bsfD{\bm{\mathsf{D}}}
\def\bsfE{\bm{\mathsf{E}}}
\def\bsfF{\bm{\mathsf{F}}}
\def\bsfG{\bm{\mathsf{G}}}
\def\bsfH{\bm{\mathsf{H}}}
\def\bsfI{\bm{\mathsf{I}}}
\def\bsfJ{\bm{\mathsf{J}}}
\def\bsfK{\bm{\mathsf{K}}}
\def\bsfL{\bm{\mathsf{L}}}
\def\bsfM{\bm{\mathsf{M}}}
\def\bsfN{\bm{\mathsf{N}}}
\def\bsfO{\bm{\mathsf{O}}}
\def\bsfP{\bm{\mathsf{P}}}
\def\bsfQ{\bm{\mathsf{Q}}}
\def\bsfR{\bm{\mathsf{R}}}
\def\bsfS{\bm{\mathsf{S}}}
\def\bsfT{\bm{\mathsf{T}}}
\def\bsfU{\bm{\mathsf{U}}}
\def\bsfV{\bm{\mathsf{V}}}
\def\bsfW{\bm{\mathsf{W}}}
\def\bsfX{\bm{\mathsf{X}}}
\def\bsfY{\bm{\mathsf{Y}}}
\def\bsfZ{\bm{\mathsf{Z}}}

% sans serif lowercase
\def\bydef{:=}
\def\sfa{{\mathsf{a}}}
\def\sfb{{\mathsf{b}}}
\def\sfc{{\mathsf{c}}}
\def\sfd{{\mathsf{d}}}
\def\sfee{{\mathsf{e}}}
\def\sfff{{\mathsf{f}}}
\def\sfg{{\mathsf{g}}}
\def\sfh{{\mathsf{h}}}
\def\sfi{{\mathsf{i}}}
\def\sfj{{\mathsf{j}}}
\def\sfk{{\mathsf{k}}}
\def\sfl{{\mathsf{l}}}
\def\sfm{{\mathsf{m}}}
\def\sfn{{\mathsf{n}}}
\def\sfo{{\mathsf{o}}}
\def\sfp{{\mathsf{p}}}
\def\sfq{{\mathsf{q}}}
\def\sfr{{\mathsf{r}}}
\def\sfs{{\mathsf{s}}}
\def\sft{{\mathsf{t}}}
\def\sfu{{\mathsf{u}}}
\def\sfv{{\mathsf{v}}}
\def\sfw{{\mathsf{w}}}
\def\sfx{{\mathsf{x}}}
\def\sfy{{\mathsf{y}}}
\def\sfz{{\mathsf{z}}}
\def\sf0{{\mathsf{0}}}

% bold sans serif lowercase
\def\bsfa{{\bm{\mathsf{a}}}}
\def\bsfb{{\bm{\mathsf{b}}}}
\def\bsfc{{\bm{\mathsf{c}}}}
\def\bsfd{{\bm{\mathsf{d}}}}
\def\bsfee{{\bm{\mathsf{e}}}}
\def\bsfff{{\bm{\mathsf{f}}}}
\def\bsfg{{\bm{\mathsf{g}}}}
\def\bsfh{{\bm{\mathsf{h}}}}
\def\bsfi{{\bm{\mathsf{i}}}}
\def\bsfj{{\bm{\mathsf{j}}}}
\def\bsfk{{\bm{\mathsf{k}}}}
\def\bsfl{{\bm{\mathsf{l}}}}
\def\bsfm{{\bm{\mathsf{m}}}}
\def\bsfn{{\bm{\mathsf{n}}}}
\def\bsfo{{\bm{\mathsf{o}}}}
\def\bsfp{{\bm{\mathsf{p}}}}
\def\bsfq{{\bm{\mathsf{q}}}}
\def\bsfr{{\bm{\mathsf{r}}}}
\def\bsfs{{\bm{\mathsf{s}}}}
\def\bsft{{\bm{\mathsf{t}}}}
\def\bsfu{{\bm{\mathsf{u}}}}
\def\bsfv{{\bm{\mathsf{v}}}}
\def\bsfw{{\bm{\mathsf{w}}}}
\def\bsfx{{\bm{\mathsf{x}}}}
\def\bsfy{{\bm{\mathsf{y}}}}
\def\bsfz{{\bm{\mathsf{z}}}}
\def\bsf0{{\bm{\mathsf{0}}}}

% Mathematical symbols
\newcommand{\Exp}[1]{{\mathbb{E}\left\{#1\right\}}}     % Expectation
\newcommand{\diag}[1]{{\mathrm{diag}\left\{#1\right\}}} % Diag operator
\newcommand{\trace}[1]{{\mathrm{Tr}\left\{#1\right\}}}  % Trace operator
\newcommand{\norm}[1]{\left\lVert#1\right\rVert}        % Norm

\newcommand{\rmF}{\mathrm{F}}               % Frobenius norm
\newcommand{\rmH}{\mathrm{H}}               % Hermitian
\newcommand{\rmT}{\mathrm{T}}               % Transpose

% System parameters
\newcommand{\NT}{N_{\rm T}}                         % Number of transmit antennas
\newcommand{\NR}{N_{\rm R}}                         % Number of receiver antennas
\newcommand{\Nu}{N_{\rm u}}                         % Number of ue antennas
\newcommand{\NRF}{N_{\rm RF}}                       % Number of RF chains

\newcommand{\Pt}{P_{\rm t}}                         % Total downlink transmit power
\newcommand{\Pmax}{P_{\rm max}}                         % Total downlink transmit power
\newcommand{\Ptu}{\tilde{P}_{{\rm t},\tu}}          % Transmit power of uplink user \tilde{u}
\newcommand{\Ptj}{\tilde{P}_{{\rm t},j}}            % Transmit power of uplink user j

\newcommand{\vardl}{\sigma_{\u}^2}                   % Noise variance at downlink user u
\newcommand{\barvardl}{\bar{\sigma}_{\u}^2}          % Interference plus noise variance at downlink user u
\newcommand{\varul}{\tilde{\sigma}^2}               % Noise variance at the BS

\newcommand{\deltaf}{\Delta f}                      % Subcarrier spacing
\newcommand{\Ts}{T_{\rm s}}                         % Symbol duration
\newcommand{\fc}{f_{\rm c}}                         % Carrier frequency
\newcommand{\lambdac}{\lambda_{\rm c}}              % Wavelength at carrier frequency
\newcommand{\lambdam}{\lambda_{m}}                  % Wavelength at m-th subcarrier

% Array steering vectors
\newcommand{\aT}{\ba_{\rm T}}                       % Array steering vector for the TX array
\newcommand{\aR}{\ba_{\rm R}}                       % Array steering vector for the RX array
\newcommand{\au}{\ba_{\rm u}}                       % Array steering vector for the UE arrays

\newcommand{\AT}{\bA_{\rm T}}                       % Transmit covariance
\newcommand{\AR}{\bA_{\rm R}}                       % Receive covariance

% Precoders and combiners
\newcommand{\FRF}{\bF_{\mathrm{RF}}}                % Analog precoder
\newcommand{\fBB}{\bff_{\mathrm{BB},1}}           % Digital precoder of user u at the m-th subcarrier
\newcommand{\fBBx}{\bff_{\mathrm{BB},\u}}          % Digital precoder of user i 
\newcommand{\fBBK}{\bff_{\mathrm{BB},\K}}          % Digital precoder of user K
\newcommand{\FBB}{\bF_{\mathrm{BB}}}              % Overall digital precoder at the m-th subcarrier
\newcommand{\FHYB}{\bF}
\newcommand{\udl}{\bu_{u,m}}                        % Digital combiner of user u at the m-th subcarrier

\newcommand{\tu}{\tilde{u}}                         % Uplink user index
\newcommand{\K}{U}                         % Number of users

\newcommand{\N}{N}                         % Number of sats

\newcommand{\WRF}{\bW_{\mathrm{RF}}}                % Analog precoder
\newcommand{\tWRF}{\tilde{\bW}_{\mathrm{RF}}}                % Analog precoder
\newcommand{\wBB}{\bw_{\mathrm{BB},\u}}          % Digital combiner of user \tilde{u} at the m-th subcarrier

\newcommand{\twBB}{\tilde{\bw}_{\mathrm{BB},\tu,m}}          % Digital combiner of user \tilde{u} at the m-th subcarrier
\newcommand{\wBBj}{\bw_{\mathrm{BB},j,m}}           % Digital combiner of user j at the m-th subcarrier
\newcommand{\WBB}{\bW_{\mathrm{BB},m}}              % Overall digital combiner at the m-th subcarrier
\newcommand{\tWBB}{\tilde{\bW}_{\mathrm{BB},m}}              % Overall digital combiner at the m-th subcarrier

\newcommand{\vul}{\bv_{\tu,m}}                      % Digital precoder at user \tilde{u} at the m-th subcarrier
\newcommand{\vulj}{\bv_{j,m}}                       % Digital precoder at user j at the m-th subcarrier

\newcommand{\wrad}{\bw_{\mathrm{t},m}}              % Radar combiner at the m-th subcarrier

% Downlink channel
\newcommand{\hSat}{\bh_i}
\newcommand{\hSati}{\bh_1}
\newcommand{\hSatN}{\bh_{\N}}

\newcommand{\HSat}{\bH_{\rm Sat}}
\newcommand{\Hdl}{\bH_{{\rm dl},u,m}}               % Downlink channel of user u
\newcommand{\barhdl}{\bar{\bh}_{{\rm dl},u,m}}      % Combined downlink channel of user u
\newcommand{\barHdl}{\bar{\bH}_{{\rm dl},m}}        % Overal combined downlink channel

\newcommand{\Lu}{L_p}                               % Number of paths
\newcommand{\alphal}{\alpha_{\ell,\u}}                      % Complex gain of l-th path
\newcommand{\taul}{\tau_\ell}                             % Delay of l-th path
\newcommand{\thetal}{\theta_{\ell,\u}}                      % AoD of l-th path user k
\newcommand{\phil}{\phi_{\ell,\u}}                          % AoD of l-th path user k
\newcommand{\tthetal}{\tilde\theta_{i}}                      % AoD of l-th path user k
\newcommand{\tphil}{\tilde\phi_{i}}                          % AoD of l-th path user k

%  channel
% \newcommand{\Hul}{\bH_{\u}}             %  channel of user \tilde{u}
% \newcommand{\Hulj}{\bH_{\j}}             % Uplink channel of user j
% \newcommand{\barhul}{\bar{\bh}_{{\rm ul},\tu,m}}    % Precoded uplink channel of user \tilde{u}
% \newcommand{\barHul}{\bar{\bH}_{{\rm ul},m}}        % Overall uplink channel
% \newcommand{\barhulj}{\bar{\bh}_{{\rm ul},j,m}}     % Precoded uplink channel of user j
\newcommand{\Hk}{\bH_{\u}}             % Uplink channel of user j
\newcommand{\Sk}{\textsf{SINR}_{\u}}
\newcommand{\ej}{\mathbf{e}_j}
\newcommand{\ek}{\mathbf{e}_{\u}}
% Target channel
\newcommand{\Ht}{\bA_{m,n}}                         % Target channel

\newcommand{\xik}{\xi_k}                            % Complex gain of the k-th target
\newcommand{\varthetak}{\vartheta_k}                % Angle of the k-th target
\newcommand{\tauk}{t_k}                          % Delay of the k-th target
\newcommand{\fDk}{f_{{\rm D},k}}                    % Doppler shift of the k-th target

% SI channel
\newcommand{\HSI}{\bH_{{\rm SI},m}}                 % SI channel

% CCI channel
\newcommand{\Hc}{\bH_{{\rm c},u,\tu,m}}             % Cross-channel interference from UL user \tilde_{u} to DL user u
\newcommand{\hc}{h_{{\rm c},u,\tu,m}}               % Effective cross-channel interference from UL user \tilde_{u} to DL user u

% Received signals
\newcommand{\ydlu}{y_{\u}}                       % Received signal at the k-th user
\newcommand{\ndl}{\bz_{\u}}                      % Noise vector at u-th user
\newcommand{\sdl}{\bd_{m,n}}                        % DL data symbols
\newcommand{\sdlu}{d_{u,m,n}}                       % DL data symbols of user u
\newcommand{\sdli}{d_{i,m,n}}                       % DL data symbols of user u

\newcommand{\yul}{\by_{m,n}}                % Received signal at the BS after RF chains
\newcommand{\nul}{\bn_{m,n}}                % Noise vector at the BS
\newcommand{\yulu}{y_{\tu,m,n}}             % Received signal at the BS for the user \tilde{u}
\newcommand{\sul}{\bs_{m,n}}                % DL data symbols
\newcommand{\sulu}{s_{\tu,m,n}}             % DL data symbols
\newcommand{\sulj}{s_{j,m,n}}               % DL data symbols

% Performance metrics and objective/cost functions
\newcommand{\Rdlu}{\mathcal{R}_{u,m}}               % Downlink spectral efficiency of user u at m-th subcarrier
\newcommand{\Rdl}{\mathcal{R}_{m}}                  % Total downlink spectral efficiency at m-th subcarrier
\newcommand{\Rulu}{\tilde{\mathcal{R}}_{\tu,m}}     % Uplink spectral efficiency of user \tilde{u} at m-th subcarrier
\newcommand{\Rulup}{\tilde{\mathcal{R}}'_{\tu,m}}   % Uplink spectral efficiency of user \tilde{u} at m-th subcarrier
\newcommand{\Rul}{\tilde{\mathcal{R}}_{m}}          % Total uplink spectral efficiency at m-th subcarrier
\newcommand{\Rulp}{\tilde{\mathcal{R}}'_{m}}        % Total uplink spectral efficiency at m-th subcarrier

\newcommand{\gaindl}[1]{\mathcal{G}_{m}(#1)}        % Downlink gain at a certain angle at m-th subcarrier
\newcommand{\gainul}[1]{\tilde{\mathcal{G}}(#1)}    % Downlink gain at a certain angle at m-th subcarrier

\newcommand{\Bdl}{\mathcal{B}_{m}}                  % Beampattern error TX
\newcommand{\Bul}{\tilde{\mathcal{B}}}              % Beampattern error RX

\newcommand{\SIpm}{\mathcal{S}_{p,m}}               % SI at the p-th receive antenna
\newcommand{\SIm}{\mathcal{S}_{m}}                  % Total SI at all the receive antennas

\newcommand{\mut}{\mu_{\rm t}}                      % Gain penalty term

\newcommand{\mub}{\mu_{\rm b}}                      % Gain penalty term
\newcommand{\muSI}{\mu_{\rm SI}}                    % SI penalty term

\newcommand{\SIqm}{\tilde{\mathcal{S}}_{q,m}}       % SI at the q-th RF chain
\newcommand{\SI}{{\tilde{\mathcal{S}}_{m}}}         % Total SI at all the RF chains

\newcommand{\tmub}{\tilde{\mu}_{\rm b}}                     % Gain penalty term
\newcommand{\tmuSI}{\tilde{\mu}_{\rm SI}}                   % SI penalty term

\newcommand{\lambdat}{\tilde{lambda}_{\rm t}}                      % Gain penalty term
\newcommand{\lambdaSI}{\tilde{lambda}_{\rm SI}}                    % SI penalty term
\newcommand{\lambdaSat}{\lambda_{\rm Sat}} 

\newcommand{\uk}{a_{\u}} 
\newcommand{\num}{\mathrm{num}_{\u}}
\newcommand{\den}{\mathrm{den}_{\u}}
\newcommand{\ck}{c_k}
\newcommand{\Nk}{N_k}
\newcommand{\Mkj}{M_{k,j}}
\newcommand{\xkj}{x_{k,j}}
\newcommand{\itBB}{\text{iter}_{\rm BB}}
\newcommand{\itRF}{\text{iter}_{\rm RF}}

\newcommand{\tot}{T}

\newcommand{\Fhyb}{\bF}
\newcommand{\Ryy}{\bR_{yy}}
\newcommand{\Rsig}{\bR_{\rm sig}}
\renewcommand{\u}{u}
\newcommand{\Mk}{\mathbf{M}_{\u}}
\newcommand{\Ru}{R_{\u}}